\documentclass{elsart}
\begin{document}
\begin{frontmatter}
\title{On wave propagation\\ in inhomogeneous systems}
\author[poly]{A.~Modinos},
\author[uoa]{N.~Stefanou},
\author[poly,uoa]{I.~E.~Psarobas\thanksref{EMAIL}}, and
\author[poly]{V.~Yannopapas}.
\address[poly]{Department of Physics, National Technical University
of Athens, \\ Zografou Campus, GR-157 73, Athens, Greece}
\address[uoa]{Section of Solid State Physics, University of Athens,\\
Panepistimioupolis, GR-157 84, Athens, Greece}
\thanks[EMAIL]{E-mail: ipsarob@cc.uoa.gr }

\maketitle

\begin{abstract}
We present a theory of electron, electromagnetic, and elastic wave
propagation in systems consisting of non-overlapping scatterers in a
host medium. The theory provides a framework for a unified
description of wave propagation in three-dimensional periodic
structures, finite slabs of layered structures, and systems with
impurities: isolated impurities, impurity aggregates, or randomly
distributed impurities. We point out the similarities and differences
between the different cases considered, and discuss the numerical
implementation of the formalism.
\begin{keyword}
Wave propagation; Inhomogeneous systems; Multiple scattering;
Photonic crystals; Phononic crystals
\end{keyword}
\end{abstract}
\end{frontmatter}
%%%%%%%%%%%%%%%%%%%%%%%%%%% SECTION 1 %%%%%%%%%%%%%%%%%%%%%%%%%%%%%%
\section{Introduction}
\label{intro} The Korringa-Kohn-Rostoker (KKR) method~\cite{KKR} has
been used extensively in the study of the electronic structure and
related properties of materials. It has been very successful in
calculations of the electronic structure of ordered elemental
solids~\cite{Moruzzi}, impurities~\cite{Dederichs}, and disordered
alloys~\cite{Gonis}. Based on the multiple-scattering
theory~\cite{Rayleigh}, the KKR method calculates from first
principles quantities such as the electron density and the
ground-state energy of the system. A KKR method for layered systems,
the so-called layer KKR (LKKR) method, has also been
developed~\cite{Kambe,LKKR}. This method is well adapted to
calculations of low-energy electron diffraction (LEED)~\cite{Pendry}
and electron emission~\cite{Modinos} spectra.

In recent years, propagation of classical waves [electromagnetic (EM)
or elastic waves] in composite materials with dielectric or,
respectively, elastic properties which are periodic functions of the
position, with a period comparable to the wavelength of the
corresponding field, has been the object of considerable
attention~\cite{Soukoulis}. These materials, photonic and phononic
crystals, respectively, whether they exist naturally or are
artificially fabricated, exhibit a rich variety of physical
properties of interest to fundamental and applied research. There are
striking analogies between the propagation of electrons in ordinary
crystals and EM/elastic waves in photonic/phononic crystals, so that
a great variety of multiple-scattering methods originally developed
for electronic-structure calculations have been transferred to the
field of photonic and phononic
crystals~\cite{Ohtaka,Ashcroft,Mod+Stef,VKKR,psm00}. The aim of the
present article is to present the KKR formalism for electron, EM and
elastic waves from a unifying point of view.
%%%%%%%%%%%%%%%%%%%%%%%%%%% SECTION 2 %%%%%%%%%%%%%%%%%%%%%%%%%%%%%%
\section{Multipole expansion of a wave field}
\label{multexp} We first consider, as an example of a field with one
degree of freedom, the case of a spinless particle of mass $m$ and
energy $\mathcal{E}$, in a potential $V({\bf r})$. According to
quantum mechanics, the particle is described by a scalar field: the
single-particle wave function
\begin{equation}
\Psi ({\bf r};t)=\Psi ({\bf r})\exp \left(-i \mathcal{E}
t/\hbar\right)\,, \label{eq:scwf}
\end{equation}
where $\hbar$ is the Planck constant and $\Psi ({\bf r})$ satisfies
the time-independent Schr\"{o}dinger equation~\cite{Bohm}
\begin{equation}
\left[-\frac{\hbar^2}{2m}\nabla^2+V({\bf r})-\mathcal{E}\right] \Psi
=0\,. \label{eq:Schrodinger}
\end{equation}
In a region of constant potential, $V_0$, Eq.~(\ref{eq:Schrodinger})
reduces to the Helmholtz wave equation
\begin{equation}
\left[\nabla^2+q^2\right] \Psi =0\,, \label{eq:Helmholtz}
\end{equation}
where $q=\sqrt{2m(\mathcal{E}-V_0)}/\hbar$. A complete set of
spherical-wave solutions of Eq.~(\ref{eq:Helmholtz}) is given by
\begin{equation}
\Psi_{\ell m}({\bf r})=f_{\ell}(qr)Y_{\ell}^{m}({\bf \hat r})\,,
\label{eq:multscal}
\end{equation}
where $Y_{\ell}^{m}({\bf \hat r})$ are the usual spherical harmonics
and $f_{\ell}$ may be any linear combination of the spherical Bessel
function, $j_{\ell}$, and the spherical Hankel function,
$h_{\ell}^{+}$. The most general wave function in a constant
potential field can be written as linear sum of the spherical waves
given by Eq.~(\ref{eq:multscal}), as follows:
\begin{equation}
\Psi({\bf r})=\sum_{\ell m} {\rm a}_{\ell
m}f_{\ell}(qr)Y_{\ell}^{m}({\bf \hat r})\,, \label{eq:scalsol}
\end{equation}
where $ {\rm a}_{\ell m}$ are coefficients to be determined.

A vector field has in general, not one, but three degrees of freedom.
Let us, for example, consider a harmonic elastic wave, of angular
frequency $\omega$, propagating in a homogeneous medium of mass
density $\rho$ and Lam\'e coefficients $\lambda$, $\mu$. This is
described by a displacement vector field:
\begin{equation}
{\bf U}({\bf r},t)={\rm Re}\left[{\bf U}({\bf r}){\rm exp}(-{\rm
i}\omega t)\right]\,, \label{eq:harmonicelastic}
\end{equation}
where ${\bf U}({\bf r})$ satisfies the equation~\cite{Landau}
\begin{equation}
\left(\lambda+2\mu\right)\nabla\left(\nabla\cdot {\bf U}\right)-\mu
\nabla\times\left(\nabla\times {\bf U} \right)+\rho \omega^2{\bf
U}=0\,. \label{eq:eltindep}
\end{equation}
The most general solution of Eq.~(\ref{eq:eltindep}) consists of two
types of elastic waves which propagate independently. These are
purely longitudinal (irrotational) waves ($\nabla\times{\bf U}={\bf
0}$), which satisfy the vector Helmholtz equation
\begin{equation}
\left[\nabla^2+q_l^2\right]{\bf U}=0\,, \label{eq:acoustic}
\end{equation}
where $q_l=\omega/c_l$, with $c_l=\sqrt{(\lambda+2\mu)/\rho}$ being
the longitudinal-wave speed; and purely transverse (divergence-free)
waves ($\nabla\cdot{\bf U}=0$), which satisfy the vector Helmholtz
equation
\begin{equation}
\left[\nabla^2+q_t^2\right]{\bf U}=0\,, \label{eq:em}
\end{equation}
where $q_t=\omega/c_t$, with $c_t=\sqrt{\mu/\rho}$ being the
transverse-wave speed. A complete set of longitudinal ($L$)
spherical-wave solutions of Eq.~(\ref{eq:acoustic}) is given
by~\cite{Chew}
\begin{equation}
{\bf U}_{\ell m}^L({\bf r})=\frac{1}{q_l} \nabla
\left[f_{\ell}(q_lr)Y_{\ell}^m({\bf \hat r})\right]\,. \label{eq:UL}
\end{equation}
A complete set of transverse ($M,N$) spherical-wave solutions of
Eq.~(\ref{eq:em}) is given by~\cite{Chew}
\begin{equation}
{\bf U}_{\ell m}^M({\bf r})=f_{\ell}(q_tr){\bf X} _{\ell m}({\bf \hat
r}) \label{eq:UM}
\end{equation}
and
\begin{equation} {\bf U}_{\ell m}^N({\bf r})=\frac{{\rm
i}}{q_t}\nabla\times f_{\ell}(q_tr){\bf X} _{\ell m}({\bf \hat r})\,,
\label{eq:UN}
\end{equation}
where ${\bf X} _{\ell m}({\bf \hat r})$ are the so-called vector
spherical harmonics, which are defined by
\begin{equation}
\sqrt{\ell (\ell+1)}{\bf X}_{\ell m}({\bf \hat r})={\bf
L}Y_{\ell}^m({\bf \hat r})\equiv -{\rm i}{\bf r}\times\nabla
Y_{\ell}^m({\bf \hat r})\,. \label{eq:Xlm}
\end{equation}
The most general displacement field in a homogeneous medium can be
written as a linear sum of the spherical waves given by
Eqs.~(\ref{eq:UL})-(\ref{eq:UN}), as follows
\begin{eqnarray}
{\bf U}({\bf r})= & & \sum_{\ell m} \biggl\{{\rm a}^M_{\ell m}
f_{\ell}(q_tr){\bf X} _{\ell m}({\bf \hat r})+{\rm a}^N_{\ell
m}\frac{{\rm i}}{q_t}\nabla\times f_{\ell}(q_tr){\bf X} _{\ell
m}({\bf \hat r})\nonumber \\ & & +{\rm a}^L_{\ell m} \frac{1}{q_l}
\nabla \left[f_{\ell}(q_lr)Y_{\ell}^m({\bf \hat r})\right]\biggr\}\;,
\label{eq:multel}
\end{eqnarray}
where ${\rm a}^P_{\ell m}$, $P=M,N,L$, are coefficients to be
determined.

Another example of a vector field is the EM field. A harmonic EM
wave, of angular frequency $\omega$, is described by its
electric-field component
\begin{equation}
{\bf E}({\bf r},t)={\rm Re}\left[{\bf E}({\bf r}){\rm exp}(-{\rm
i}\omega t)\right]\,. \label{eq:harmonicelectric}
\end{equation}
We need not write down explicitly the magnetic-field component of the
wave since this can be readily obtained from ${\bf E}({\bf r},t)$
using one of Maxwell equations. In a homogeneous medium characterized
by a dielectric function $\epsilon(\omega)\epsilon_0$ and a magnetic
permeability $\mu(\omega)\mu_0$, where $\epsilon_0$, $\mu_0$ are the
electric permittivity and magnetic permeability of vacuum, Maxwell
equations imply that ${\bf E}({\bf r})$ satisfies a vector Helmholtz
equation, subject to the condition $\nabla \cdot {\bf E}=0$, with a
wave number $q=\omega/c$, where $c=1/\sqrt{\mu \epsilon \mu_0
\epsilon_0}$ is the velocity of light in the medium. The
spherical-wave expansion of ${\bf E}({\bf r})$ is given by the terms
of Eq.~(\ref{eq:multel}) which satisfy $\nabla \cdot {\bf E}=0$,
i.e.,
\begin{eqnarray}
{\bf E}({\bf r})=\sum_{\ell m} \left\{{\rm a}^H_{\ell m}
f_{\ell}(qr){\bf X} _{\ell m}({\bf \hat r})+{\rm a}^E_{\ell
m}\frac{{\rm i}}{q}\nabla\times \left[f_{\ell}(qr){\bf X} _{\ell
m}({\bf \hat r})\right]\right\}\;, \label{eq:multem}
\end{eqnarray}
where ${\rm a}^P_{\ell m}$, $P=H,E$, are coefficients to be
determined.
%%%%%%%%%%%%%%%%%%%%%%%%%%% SECTION 3 %%%%%%%%%%%%%%%%%%%%%%%%%%%%%%
\section{Scattering by a single scatterer}
\label{sphscat} We consider a scatterer of finite range, $S$, with
its center at the origin of coordinates, and assume that the
appropriate in each case characteristics (potential, Lam\'e
coefficients and mass density, electric permittivity and magnetic
permeability) are different from those of the surrounding homogeneous
medium. A De Broglie, EM or an elastic plane wave incident on this
scatterer is described, respectively, by Eq.~(\ref{eq:multscal}),
(\ref{eq:multem}) or (\ref{eq:multel}) with $f_{\ell}=j_{\ell}$
(since the plane wave is finite everywhere) and appropriate
coefficients ${\rm a}_L^0$, where $L$ denotes collectively the
indices $\ell m P$. Similarly, the scattered wave is described by
Eq.~(\ref{eq:multscal}), (\ref{eq:multem}) or (\ref{eq:multel}) with
$f_{\ell}=h^+_{\ell}$, which has the asymptotic form appropriate to
an outgoing spherical wave: $h^+_{\ell}\approx (-{\rm
i})^{\ell}\exp({\rm i} q r)/{\rm i} q r$ as $r\rightarrow\infty$, and
appropriate expansion coefficients ${\rm a}^+_L$. The wavefield for
$r>S$ is the sum of the incident and scattered waves. The
spherical-wave expansion of the field for $r<S$ is obtained in
similar manner by the requirement that it be finite at the origin
(${\bf r}={\bf 0}$). By applying the proper in each case boundary
conditions at the surface of the scatterer, we obtain a relation
between the expansion coefficients of the incident and the scattered
field, as follows:
\begin{equation}
{\rm a}^+_L=\sum_{L'} T_{LL'}\, {\rm a}^0_{L'}\;, \label{eq:tmatrix}
\end{equation}
where $T_{LL'}$ are the elements of the so-called scattering
transition matrix. Eq.~(\ref{eq:tmatrix}) is valid for any shape of
scatterer; for spherically symmetric scatterers each spherical wave
scatters independently of all others, which leads to a transition
matrix diagonal in angular momentum. Explicit forms for $T_{LL'}$ for
various cases can be found elsewhere~\cite{Bohm,Brill,Bohren}.
%%%%%%%%%%%%%%%%%%%%%%%%%%% SECTION 4 %%%%%%%%%%%%%%%%%%%%%%%%%%%%%%
\section{Multiple scattering theory}
\label{multscat} We consider an assembly of non-overlapping
scatterers centered on sites ${\bf R}_i$ in a homogeneous medium. In
general: an outgoing wave about ${\bf R}_i'$ [described by
Eq.~(\ref{eq:multscal}), (\ref{eq:multel}) or (\ref{eq:multem}) with
$f_{\ell}=h^+_{\ell}$ and expansion coefficients ${\rm b}^{+i'}_L$]
can be expanded into spherical waves about ${\bf R}_i$, incident on
${\bf R}_i$. This expansion has the form of Eq.~(\ref{eq:multscal}),
(\ref{eq:multel}) or (\ref{eq:multem}) with $f_{\ell}=j_{\ell}$ and
expansion coefficients ${{\rm b}'}_L^i(i')$, given by
\begin{equation}
{{\rm b}'}_L^i(i')=\sum_{L'} \Omega_{LL'}^{\,ii'}\,{\rm
b}_{L'}^{+i'}\;. \label{eq:expcoef}
\end{equation}
Explicit expressions for the so-called free-space structural Green
functions, $\Omega_{LL'}^{\,ii'}$, for the scalar and transverse
vector fields can be found elsewhere~\cite{Gonis,SM93}. In the case
of the elastic field, since a longitudinal (transverse) spherical
wave about ${\bf R}_{i'}$ remains a longitudinal (transverse) wave
when expanded about another center ${\bf R}_{i}\neq{\bf R}_{i'}$, the
matrix elements $\Omega_{LL'}^{\,ii'}$ are obtained independently for
longitudinal and transverse waves. Therefore, $\Omega_{LL'}^{\,ii'}$
has a block-diagonal form in the mode index $P$ (we recall that
$L\equiv\ell m P$): the matrix elements of the block corresponding to
transverse waves are identical to the corresponding matrix elements
for the EM field, while the matrix elements of the block
corresponding to longitudinal waves are identical to the
corresponding matrix elements for the scalar field~\cite{psm00}. We
note that, for any field, by definition, $\Omega_{LL'}^{\,ii'}$
equals zero for $i=i'$.

The wave scattered from the scatterer at ${\bf R}_i$ is determined
from the total wave incident on this scatterer; therefore
\begin {equation}
{\rm b}^{+i}_L=\sum_{L'} T_{LL'}^i\,({\rm a}^{0i}_{L'}+{{\rm
b}'}_{L'}^i)\;, \label{eq:bTab}
\end{equation}
where ${\rm a}^{0i}_{L}$ are the coefficients in the multipole
expansion [given by Eq.~(\ref{eq:multscal}), (\ref{eq:multel}) or
(\ref{eq:multem})] about ${\bf R}_i$ of an externally incident wave;
and ${{\rm b}'}_{L}^i$ are the coefficients in the multipole
expansion about ${\bf R}_i$ of the outgoing waves scattered from all
the other scatterers at ${\bf R}_{i'}\neq{\bf R}_i$. From
Eq.~(\ref{eq:expcoef}) we have
\begin{equation}
{{\rm b}'}_{L}^i=\sum_{i',L'} \Omega_{LL'}^{\,ii'}\,{\rm
b}_{L'}^{+i'}\;. \label{eq:bob}
\end{equation}
Substituting Eq.~(\ref{eq:bob}) into Eq.~(\ref{eq:bTab}) we obtain
\begin{equation}
\sum_{i',L'}\left(\delta_{ii'}\delta_{LL'}-\sum_{L''}T_{LL''}^i
\,\Omega_{L''L'}^{\,ii'}\right){\rm
b}_{L'}^{+i'}=\sum_{L'}T_{LL'}^i\,{\rm a}^{0i}_{L'}\;.
\label{eq:modeq}
\end{equation}

We now introduce the structural Green functions $D_{LL'}^{\,ii'}$
which give the coefficients [in an expansion such as
(\ref{eq:multscal}), (\ref{eq:multel}) or (\ref{eq:multem})] of the
wave incident on the scatterer at ${\bf R}_i$, due to an outgoing
wave from the scatterer at ${\bf R}_{i'}$. An outgoing wave from the
$i'$th scatterer can reach the $i$th scatterer directly, or
indirectly after scattering any number of times by any number of
scatterers (including those at ${\bf R}_i$ and ${\bf R}_{i'}$). Let
$D_{LL'}^{\,ii'}$ express the sum of the contribution to the
coefficients of the incident wave on ${\bf R}_i$ from all possible
scattering paths originating from an outgoing wave from the $i'$th
scatterer. One can easily prove by iteration the following equation
\begin{equation}
D_{LL'}^{\,ii'}=\Omega_{LL'}^{\,ii'}+\sum_{i'',L'',L'''}
\Omega_{LL''}^{\,ii''} T_{L''L'''}^{\,i''}D_{L'''L'}^{\,i''i'}\;,
\label{eq:DLi}
\end{equation}
where $T_{L''L'''}^{\,i''}$ are the elements of the transition matrix
which describes the scattering by the single scatterer at ${\bf
R}_{i''}$. We generalize Eq.~(\ref{eq:DLi}) by treating the
scattering at ${\bf R}_i$ in two stages. The first-stage scattering
is described by $T_{LL'}^{ri}$, which correspond to arbitrarily
defined scatterers (reference scatterers), and the second-stage
scattering by $T_{LL'}^{\,i}-T_{LL'}^{ri}$. We obtain:
\begin{equation}
D_{LL'}^{\,ii'}=D_{LL'}^{r\,ii'}+\sum_{i'',L'',L'''}D_{LL''}^{r\,ii''}
\left(T_{L''L'''}^{\,i''}-T_{L''L'''}^{ri''}\right)D_{L'''L'}^{\,i''i'}\;,
\label{eq:D2}
\end{equation}
where $D_{LL'}^{r\,ii'}$ are the solution of Eq.~(\ref{eq:DLi}) when
$T_{LL'}^{\,i}=T_{LL'}^{ri}$.
%%%%%%%%%%%%%%%%%%%%%%%%%%% SECTION 5 %%%%%%%%%%%%%%%%%%%%%%%%%%%%%%
\section{Bulk systems}
\label{bulk} We consider the case in which the scatterers constitute
a three dimensional (3D) crystal structure specified by Bravais
lattice vectors ${\bf R}_n$ and non-primitive translation vectors,
indicating the positions of the scatterers within the unit cell,
${\bf t}_{\alpha}$; in this case the site index $i$ stands for the
composite index $n\alpha$. Bloch's theorem implies that ${\rm
b}_{L'}^{+n'\alpha'}=\exp\left[{\rm i} {\bf k}\cdot({\bf R}_{n'}-{\bf
R}_n)\right] {\rm b}_{L'}^{+n\alpha'}$.

The normal modes of the crystal are obtained, by putting the external
incident wave equal to zero in Eq.~(\ref{eq:modeq}), which leads to
the following secular equation:
\begin{equation}
\det\left[\delta_{\alpha
\alpha'}\delta_{LL'}-\sum_{L''}T_{LL''}^{\alpha}
\,\Omega_{L''L'}^{\,\alpha \alpha'}({\bf k})\right]=0\,,
\label{eq:seculeq2}
\end{equation}
where
\begin{equation}
\Omega_{LL'}^{\,\alpha \alpha'}({\bf k})=\sum_{n'} \Omega_{LL'}^{\,n
\alpha;n'\alpha'} \exp\left[{\rm i} {\bf k}\cdot({\bf R}_{n'}-{\bf
R}_n)\right]\,, \label{eq:OLa}
\end{equation}
which does not depend on $n$. Both $T_{LL'}^{\alpha}$ and
$\Omega_{LL'}^{\,\alpha \alpha'}({\bf k})$ in Eq.~(\ref{eq:seculeq2})
are functions of the frequency of the wave, but the
$T_{LL'}^{\alpha}$ depend only on the properties of a single
scatterer, whereas $\Omega_{LL'}^{\,\alpha \alpha'}({\bf k})$ depend
only on the geometry. Thus Eq.~(\ref{eq:seculeq2}) reflects a
complete separation of the individual scatterer and structural
aspects of the problem. The matrix elements $\Omega_{LL'}^{\,\alpha
\alpha'}({\bf k})$ for the scalar field were introduced by Korringa,
Kohn, and Rostoker~\cite{KKR} in relation to the calculation of the
electronic band structure of periodic solids and are commonly
referred to as the KKR structure constants. The calculation of the
structure constants, which needs to be done only once for a given
lattice, usually requires Ewald-summation techniques~\cite{3D}. We
note that Eq.~(\ref{eq:seculeq2}) involves in principle
infinite-dimensional matrices. In actual calculations, however, it is
sufficient to truncate the angular momentum index, $\ell$, to some
relatively small number, $\ell_{max}$. The KKR method has been
successfully used for many years in the calculation of the electron
band structure of solids~\cite{Moruzzi}, and more recently it has
also been applied to the calculation of the frequency band structure
of EM and elastic fields in relation to photonic and phononic
crystals~\cite{VKKR}.

One of the advantages of the KKR theory compared to other approaches
is its capability of dealing with defects and disorder~\cite{Gonis}.
For the description of point defects at a finite number of sites, one
needs to calculate the structural Green functions, $D_{LL'}^{\,ii'}$,
of the defect system. This can be done in real space, using
Eq.~(\ref{eq:D2}) and considering the periodic crystal (without any
defect) as the reference system. In this case, the sum over $i''$ in
Eq.~(\ref{eq:D2}) is restricted to those sites at which there are
defects and induce a perturbation $\delta T_{L''L'''}^{\,i''}=
T_{L''L'''}^{\,i''}-T_{L''L'''}^{r\,i''}$ on the $T$ matrix. For the
periodic arrangement of scatterers of the reference system, the
evaluation of $D_{LL'}^{r\,ii'}$ through Eq.~(\ref{eq:DLi}) involves
an infinite number of sites $i''$ and it is achieved by a lattice
Fourier transform of Eq.~(\ref{eq:DLi})
\begin{equation}
D_{LL'}^{r\,\alpha \alpha'}({\bf k}) =\Omega_{LL'}^{\,\alpha \alpha
'}({\bf k})+\sum_{\alpha'',L'',L'''} \Omega_{LL''}^{\,\alpha
\alpha''}({\bf k})
T_{L''L'''}^{r\,\alpha''}D_{L'''L'}^{r\,\alpha''\alpha'}({\bf k})
\label{eq:FDLi}
\end{equation}
and subsequent integration over the volume $v$ of the Brillouin zone
(BZ)
\begin{equation}
D_{LL'}^{r\,ii'}=\frac{1}{v}\;\int_{{\rm BZ}} {\rm d}^3k \;\;
\exp\left[{\rm i} {\bf k}\cdot\left({\bf R}_n-{\bf
R}_{n'}\right)\right]D_{LL'}^{r\,\alpha \alpha'}({\bf k})\,.
\label{eq:intBZ}
\end{equation}
If a number of $N$ different scatterers, described by $T$ matrices
$T^j_{LL'}$, $j=1,2,\ldots,N$, are randomly distributed on the sites
of a given crystal, the coherent potential approximation
(CPA)~\cite{Soven} can be used to treat the disordered system within
a mean-field context. The effective CPA medium consists of identical
scatterers at all sites of sublattice $\alpha$ of the crystal,
characterized by a $T$ matrix $T^{c\, \alpha}_{LL'}$. The structural
Green functions $D_{LL'}^{c\,\, n\alpha;n'\alpha'}$ of the CPA medium
are calculated by equations analogous to (\ref{eq:FDLi}) and
(\ref{eq:intBZ}). The CPA self-consistency condition demands that the
correction to the scattering due to the difference of the actual
scatterers from the CPA scatterer vanishes on the average
\begin{equation}
\sum_{j=1}^{N} c_{j;\alpha} \left[{\bf I}-\left({\bf T}^{j}-{\bf
T}^{c\, \alpha}\right){\bf D}^{c\,\,
0\alpha;0\alpha}\right]^{-1}\;\left({\bf T}^j-{\bf T}^{c\,
\alpha}\right)=0\;\;\;\; {\rm for\ every}\  \alpha\,, \label{eq:CPA}
\end{equation}
where $c_{j;\alpha}$ denotes the concentration of scatterer $j$ on
sublattice $\alpha$.
%%%%%%%%%%%%%%%%%%%%%%%%%%% SECTION 6 %%%%%%%%%%%%%%%%%%%%%%%%%%%%%%
\section{Layered systems}
\label{layer} A layered system consists of a number of planes
(layers) of scatterers with the same two dimensional (2D)
periodicity. To begin with, we consider just one layer, at $z=0$, in
which case the scatterers are centered on the sites ${\bf R}_n+{\bf
t}_{\alpha}$, where $\{{\bf R}_n\}$ constitutes a 2D Bravais lattice
and ${\bf t}_{\alpha}$ denote the positions of the scatterers within
the 2D unit cell. The 2D reciprocal vectors ${\bf g}$, and the
surface Brillouin zone (SBZ) corresponding to this lattice are
defined in the usual manner~\cite{Pendry,Modinos}. The wave vector
parallel to the plane of atoms can always be written as follows:
\begin{equation}
{\bf q}_{\|}={\bf k}_{\|}+{\bf g}'\,, \label{eq:qkg}
\end{equation}
where the reduced wave vector, ${\bf k}_{\|}$, lies in the SBZ and
${\bf g}'$ is a certain reciprocal vector. In what follows we write
the wave vector of a plane wave of given ${\bf q}_{\|}={\bf
k}_{\|}+{\bf g}$ and given wave number $q_{\nu}$, where $\nu$
specifies the polarization mode of the wave, if there exist
longitudinal and transverse modes, as follows:
\begin{equation}
{\bf K}_{{\bf g}\nu}^{\pm}= \left({\bf k}_{\|}+{\bf g}\,,\,\pm
\left[q_{\nu}^2-({\bf k}_{\|}+{\bf g})^2\right]^{1/2}\right)\,.
\label{eq:Kg}
\end{equation}
We note that when $q_{\nu}^2<({\bf k}_{\|}+{\bf g})^2$, the above
defines a decaying wave; the positive (negative) sign in
Eq.~(\ref{eq:Kg}) corresponds to a wave propagating or decaying to
the right (left).

The coefficients ${\rm b}_L^{+n\alpha}$ in the multipole expansion of
the scattered field from the layer are obtained from
Eq.~(\ref{eq:modeq}). Using the Bloch condition: ${\rm b}_{L'}^{+n'
\alpha'}=\exp\left[{\rm i}{\bf k}_{\|}\cdot({\bf R}_{n'}-{\bf
R}_n)\right]{\rm b}_{L'}^{+n\alpha'}$, we obtain
\begin{equation}
\sum_{\alpha',L'}\left[\delta_{\alpha
\alpha'}\delta_{LL'}-\sum_{L''}T_{LL''}^{\alpha}
\,\Omega_{L''L'}^{\,\alpha \alpha'}({\bf k}_{\|})\right]{\rm
b}_{L'}^{+n \alpha'}=\sum_{L'}T_{LL'}^{\alpha}\,{\rm a}^{0 n \alpha
}_{L'}\;, \label{eq:2Dmodeq}
\end{equation}
where ${\rm a}^{0 n \alpha }_{L}$ are the coefficients in the
multipole expansion of the incident plane wave. The 2D KKR structure
constants
\begin{equation}
\Omega_{LL'}^{\,\alpha \alpha'}({\bf
k}_{\|})=\sum_{n'}\Omega_{LL'}^{\,n \alpha; n' \alpha'}
\exp\left[{\rm i}{\bf k}_{\|}\cdot({\bf R}_{n'}-{\bf R}_n)\right]\,,
\label{eq:2DKKR}
\end{equation}
like their counterparts in 3D given by Eq.~(\ref{eq:OLa}), can be
evaluated using Ewald-summation techniques~\cite{Kambe}.

Writing the incident, reflected and transmitted plane waves with
respect to an origin (a scattering center) in this plane, we obtain
the amplitudes, $\left[\Phi_{\rm rf}\right]_{{\bf g}i}^{\pm}$ and
$\left[\Phi_{\rm tr}\right]_{{\bf g}i}^{\pm}$ of the reflected and
transmitted beams, respectively, in terms of the amplitudes,
$\left[\Phi_{\rm in}\right]_{{\bf g}'i}^{\pm}$, of the incident wave
as follows:
\begin{eqnarray}
\left[\Phi_{\rm rf}\right]_{{\mathbf{g}} i}^{-s} &&
=\sum_{i'}M_{{\mathbf{g}}i;{\mathbf{g}}'i'}^{-ss} \left[\Phi_{\rm
in}\right]_{{\mathbf{g}}' i'}^s\nonumber
\\ \left[\Phi_{\rm tr}\right]_{{\mathbf{g}} i}^s
&& =\sum_{i'}M_{{\mathbf{g}}i;{\mathbf{g}}'i'}^{ss} \left[\Phi_{\rm
in}\right]_{{\mathbf{g}}' i'}^s\,, \label{eq:amplitude}
\end{eqnarray}
where $s=+(-)$ specifies a wave traveling or decaying to the positive
(negative) $z$ direction, and $i$ denotes the components of the wave
field. A scalar field ($\mbox{\boldmath $\Phi$}\rightarrow \Psi$) has
just one component ($i=1$), the displacement vector of the elastic
field ($\mbox{\boldmath $\Phi$}\rightarrow{\bf U}$) has three
components ($i=1,2,3$), while the EM field ($\mbox{\boldmath
$\Phi$}\rightarrow{\bf E}$), because of its transverse nature, has
only two independent components, $i=1,2$. Explicit expressions for
the transmission/reflection matrix elements
$M_{{\mathbf{g}}i;{\mathbf{g}}'i'}^{ss'}$ for the scalar, elastic,
and EM fields can be found
elsewhere~\cite{Pendry,Modinos,psm00,comphy2}.

The transmission/reflection matrices for a slab which consists of a
stack of layers with the same 2D periodicity parallel to a given
plane are obtained from the transmission/reflection matrices of the
individual layers in the manner described in
Refs.~\cite{Pendry,Modinos,psm00,comphy2}. Knowing the
transmission/reflection matrices for the slab we can readily obtain
the transmission, reflection, and absorption coefficients of the
plane wave of Eq.~(\ref{eq:Kg}) incident on the slab. The LKKR
technique summarized in this section is particularly well adapted to
layered systems because the computation time scales linearly to the
number of layers in the slab. Moreover, because this method does not
require periodicity in the direction perpendicular to the layers, it
can easily treat heterostructures, slabs with impurity
planes~\cite{comphy2,McLaren,KMS94,PSMsub} or even a random
succession of different layers, as long as these have the same 2D
periodicity.

The LKKR technique provides also the complex band structure of an
infinite crystal, viewed as a succession of layers parallel to a
given crystallographic plane~\cite{Pendry,Modinos,psm00,comphy2}.

Point defects or a random distribution of point defects in the layers
of a slab can also be treated along the lines described in
Section~\ref{bulk}~\cite{Gonis,SM93,MYS00}.
%%%%%%%%%%%%%%%%%%%%%%%%%%% ACK       %%%%%%%%%%%%%%%%%%%%%%%%%%%%%%
\ack{I.~E.~Psarobas was partly supported by the University of
Athens, and partly by the Institute of Communication and Computer
Systems (ICCS) of the National Technical University of Athens.
V.~Yannopapas was supported by the State Scholarship Foundation
of Greece (I.K.Y.).}
%%%%%%%%%%%%%%%%%%%%%%%%%%% REFERENCES %%%%%%%%%%%%%%%%%%%%%%%%%%%%%%

\end{document}